\documentclass[Letter,useAMS,usenatbib]{mn2e}
\usepackage{graphicx}
\usepackage{hyperref}
\usepackage{lscape}

\begin{document}

\title[X-ray bursts from the AMXP IGR~J17511--3057]{Type I X-ray
  bursts and burst oscillations in the accreting millisecond X-ray
  pulsar IGR~J17511--3057}

\author[Altamirano et al.]{D. Altamirano$^1$\thanks{E-mail:d.altamirano@uva.nl}, 
A. Watts$^1$,
M. Linares$^2$,
C. B. Markwardt$^{3,4}$,\newauthor
T. Strohmayer$^3$ and
A. Patruno$^1$
\\
  $^{1}$: Astronomical Institute, ``Anton Pannekoek'', University of
  Amsterdam, Science Park 904, 1098XH, Amsterdam, The Netherlands.\\
  $^2$: Massachusetts Institute of Technology - Kavli Institute for
  Astrophysics and Space Research, Cambridge, MA 02139, USA.\\
  $^3$: Astroparticle Physics Laboratory, Mail Code 661, NASA Goddard Space Flight Center, Greenbelt, MD 20771, U.S.A. \\ 
 $^4$:Department of Astronomy,   University of Maryland, College Park, MD 20742., U.S.A. }

\pagerange{\pageref{firstpage}--\pageref{lastpage}}
\pubyear{2009}

\maketitle

\label{firstpage}

\begin{abstract}
  We report the discovery of burst oscillations at the spin frequency
  in ten thermonuclear bursts from the accreting millisecond X-ray
  pulsar (AMXP) IGR~J17511--3057.  
  The burst oscillation properties are, like those from the persistent
  AMXPs SAX J1808.4--3658 and XTE J1814--338, anomalous compared to
  burst oscillations from intermittent pulsars or non-pulsing LMXBs.
  Like SAX J1808.4--3658 they show frequency drifts in the rising
  phase rather than the tail.  There is also evidence for harmonic
  content.  
  Where IGR J17511--3057 is unusual compared to the other two
  persistent pulsars is that oscillations are not detected throughout
  all bursts.  
  As accretion rate drops the bursts get brighter and their rise/decay
  time scales become shorter, while the oscillation amplitude falls
  below the detection threshold: first in the burst peak and then also
  in the rise.  None of the bursts from IGR J17511--3057 show evidence
  for photospheric radius expansion (which might be expected to
  suppress oscillation amplitude) which allow us to set an upper limit
  to the distance of 6.9 kpc.  We discuss the implications of our
  results for models of the burst oscillation mechanism.
\end{abstract}
\begin{keywords} 
Keywords: accretion, accretion disks --- binaries: close --- stars:
  individual (IGR J17511--3057) --- stars: neutron, pulsars --- X--rays: stars
\end{keywords}

\section{Introduction}
\label{sec:intro}

It is thought that many low-mass X-ray binaries (LMXBs) contain
rapidly rotating neutron stars. However measuring the neutron star
spin has been a challenge for more than 30 years. So far we have been
able to measure the spin of only 25 out of more than 150 LMXBs
\citep[see, e.g.,][for a catalog]{Liu07} thanks to episodes of X-ray
pulsations at the neutron star spin frequency. These episodes can last
for periods as short as a few seconds or as long as days to weeks.

When pulsations occur in the persistent X-ray emission, the sources
are known as accreting millisecond X-ray pulsars (AMXPs). Today there
are 13 of these systems \citep[see][for the latest
discovery]{Altamirano10f,Markwardt10a} and their pulsations are
thought to be powered by accretion, in a scenario in which matter from
the accretion disk is channeled by the magnetic field lines onto the
magnetic pole, forming a hot spot visible in X-rays.

When pulsations occur during surface thermonuclear explosions
\citep[known as Type I X-ray bursts, see, e.g][ for a
review]{Strohmayer06}, the sources are known as nuclear-powered X-ray
pulsars (NPXPs) or burst oscillation sources. Prior to the discovery
of IGR J17511--3057 there were 14 confirmed NPXPs \citep[including
the four which are also AMXPs, and excluding those which require
confirmation, see][]{Watts08,Altamirano10c,Galloway10}, of which two
are persistent AMXPs and two are intermittent AMXPs (see below).  For
clarity, in the rest of this work we will refer to the
accretion-powered pulsations as ``pulsations'' and to the
nuclear-powered pulsations as ``burst oscillations''.

Burst oscillations presumably develop when the burning or its
aftermath becomes in some way asymmetric.  However the exact mechanism
that generates burst oscillations is not yet known, and they are not
observed in every burst even in sources that do show them.
All theoretical models proposed to date fail to explain certain
aspects of the data, and it seems increasingly likely that the
magnetic field in the surface layers (largely neglected by current
models) may play an important role.

In this regard systems that are both AMXPs and NPXPs are particularly
useful.  The magnetic field in these sources is obviously dynamically
important, since it channels accretion flow onto the stellar surface.
This allows us to gauge field strength and any effect that channeling
may have in establishing either temperature or fuel distribution
gradients in the surface layers \citep{Strohmayer06, Watts08a,
  Watts08b, Lamb09a}.

Prior to the discovery of IGR J17511--3057, there were only four
systems that were both AMXPs and NPXPs.  If one compares the duty
cycle of their pulsations, these systems can be divided into two
separate groups.
The first group contains the AMXPs SAX~J1808.4--3658
\citep{Wijnands98} and XTE~J1814--338 \citep{Markwardt03d,
  Strohmayer03a}; these systems show pulsations persistently during
all their outbursts \citep[see][respectively]{Hartman09,Papitto07}.
The second group contains the AMXPs HETE~J1900.1--2455
\citep{Kaaret06} and Aql X-1 \citep{Casella08}; pulsations from these
sources were only detected intermittently. HETE~J1900.1--2455 has been
in outburst for almost 4.5 years, but showed pulsations only
intermittently during the first few months of activity
\citep{Galloway07a,Galloway08a}.  The intermittent AMXP Aql X-1 is
particularly extreme: pulsations were detected only for
$\approx150$~sec out of the $\approx1.5$~Msec of available data
\citep{Casella08}.

These two groups of sources differ not only in their pulsation duty
cycle, but also in the characteristics of their burst oscillations.
The intermittent pulsars show burst oscillations with similar
characteristics to those observed in all non-AMXPs
\citep{Zhang98b,Galloway08,Watts09}.  Burst oscillations are detected
in only a subset of bursts.  They show no harmonic content and are
detected more often during decay than burst rise.  Frequency drifts,
which are common, are slow.  Burst oscillation amplitudes tend to
increase with energy.
The persistent pulsars, by contrast, are both rather anomalous.  They
show burst oscillations in all bursts, throughout the whole burst
(i.e. starting at the rise, continuing at the peak -- except during
photospheric-radius expansion peaks (PRE)-- and during the decay). SAX
J1808.4-3658 shows very large fast frequency drifts in the burst rise
that overshoot the spin frequency \citep{Chakrabarty03}, while XTE
J1814-338 is remarkable for its frequency stability and phase-locking
\citep{Strohmayer03a, Watts05,Watts08a}.  XTE J1814-338 has burst
oscillations whose amplitude rises with energy \citep{Watts06a}, and
the oscillations also have substantial harmonic content.

These differences support the idea that magnetic field plays an
important part in determining burst oscillation properties, either
directly or by generating composition/temperature/fuel depth gradients
in the surface layers.  Understanding how burst oscillation properties
vary between and within the groups of persistent AMXPs, intermittent
AMXPs and non-AMXPs is the first step to shedding light onto the
mechanism (or mechanisms) that produce brightness asymmetries during
X-ray bursts.  In this paper we report the discovery of burst
oscillations from the persistent AMXP IGR J17511--3057.  We present a
detailed analysis of the burst oscillation properties and discuss how
they fit into the existing population.

\subsection{IGR~J17511--3057}\label{sec:source}

IGR~J17511--3057 was first unambiguously identified by the INTEGRAL
gamma ray mission on September 12th, 2009 \citep{Baldovin09}.
However, the source was initially detected one day before by RXTE PCA
Bulge scans as a rising flux attributed to two known sources in the
same 1$^\circ$ field of view: XTE~J1751--305 and GRS~1747--312
\citep[the former is a known $\approx435$~Hz AMXP,
see][]{Markwardt02}. The first PCA pointed observation was performed
on September 12th, and revealed coherent pulsations at $\approx$245 Hz
\citep{Markwardt09}.
Quasi-simultaneously with the PCA observation, Swift observed in
the direction of the AMXP XTE~J1751--305 and did not detect any source
at its known Chandra position, or any other position in the 22 arcmin
diameter FoV \citep{Markwardt09}.
All these observations together confirmed that the source in outburst
was not the known AMXP XTE~J1751--305 but a new one.
A preliminary orbital solution yielded a pulse frequency of
244.8337(1) Hz, an orbital period of 207.4(8) minutes, and a projected
semi-major axis of 274(1) lt-ms.  The mass function was 0.00107(2)
$M_{\sun}$, giving a minimum companion mass of $\approx0.13$ $M_{\sun}$.

Swift observations at the beginning of the outburst detected
thermonuclear (Type I) X-ray bursts \citep{Bozzo09}. RXTE follow-up
observations led to the detection of burst oscillations during these
thermonuclear bursts \citep{Watts09a}.
Chandra-HETG observations were performed on September 22nd
\citep{Nowak09}, leading to a refined position ($\alpha=17^h 51^m
08^s.66$, $\delta = -30^o 57^{'} 41^{''}$). Using the new coordinates,
\citet{Riggio09} and \citet{Papitto09a} reported an improved orbital
ephemeris (using RXTE and XMM-Newton data, respectively). Very
recently, \citet{Papitto10} presented a detailed analysis of the X-ray
spectra of IGR~J17511--3057 using 70~ksec of XMM-Newton data.
Near infrared follow-up observations identified the counterpart at a
magnitude of K$_s=18.0\pm0.1$ \citep{Torres09a,Torres09b}. Very
recently, \citet{Bozzo10} reported on the Swift monitoring
observations during the outburst.

\section{OBSERVATIONS AND DATA ANALYSIS}
\label{sec:dataanalysis}

\subsection{Light curves and color diagrams}\label{sec:lcandccd}

We use data from the Rossi X-ray Timing Explorer (RXTE) Proportional
Counter Array \citep[PCA; for instrument information
see][]{Zhang93,Jahoda06}. 
There were 85 pointed observations (ObsIDs: 94041-01 and 94042-01)
containing $\approx2.5$ to $\approx30$~ksec of useful data per observation.
We use the 16-s time-resolution Standard 2 mode data to calculate
X-ray colors.  Hard and soft color are defined as the 9.7--16.0 keV /
6.0--9.7 keV and 3.5--6.0 keV / 2.0--3.5 keV count rate ratio,
respectively, and intensity as the 2.0--16.0 keV count rate.
Type I X-ray bursts were removed, background was subtracted and
deadtime corrections were made.  Colors and intensities were
normalized by those of the Crab Nebula \citep[see][and Table 2 in
\citealt{Altamirano08} for average colors of the Crab Nebula per
PCU]{Kuulkers94,Straaten03}.

\subsection{Energy spectra of the persistent emission}

For the PCA, we used the Standard 2 data of PCU 2, which was active in
all observations. The background was estimated using the PCABACKEST
V6.0 (see FTOOLS). We calculated the PCU 2 response matrix for
each observation using the FTOOLS routine PCARSP V10.1.
For the HEXTE instrument, spectra were accumulated for cluster B (as
cluster A stopped rocking in October 2006), excluding the damaged
detector and averaging both rocking directions to measure the
background spectrum.
 Dead time corrections of both source and background spectra were
 performed using HXTDEAD V6.0. The response matrices were created
 using HXTRSP V3.1.
 For both PCA and HEXTE, we filtered out data recorded during, and up
 to 30 minutes after passage through the South Atlantic Anomaly
 (SAA). We only used data when the pointing offset from the source was
 less than 0.02 degrees and the elevation of the source with respect
 to the Earth was greater than 10 degrees. Using XSPEC V11.3.2i
 \citep{Arnaud96}, we fitted simultaneously the PCA and HEXTE energy
 spectra using the 3.0--25.0~keV and 20.0--200.0~keV energy bands,
 respectively. We used a model consisting of a disk blackbody and a
 power law, absorbed with an equivalent Hydrogen column density of
 1.1$\times$10$^{22}$ cm$^{-2}$ \citep{Papitto09a,Papitto10}.

\subsection{Type I X-ray bursts}\label{sec:resonbo}

We searched the Swift X-ray telescope (XRT) and the burst alert
telescope (BAT) data for the occurrence of X-ray bursts. We found a
total of 3 bursts. Spectral analysis of these bursts have been
reported by \citet{Bozzo10}. We refer to \citet{Bozzo10} for more
information on the Swift data. Here we only report on the search for
burst oscillations.

We searched the Standard~1 mode PCA data (2--60 keV, 0.125 seconds
time resolution, no energy resolution) of all observations and found
10 Type I X-ray bursts (see Table~\ref{table:bursts}). In the
following sections we describe the spectral and timing procedures
separately.

\subsubsection{Time resolved spectroscopy of bursts detected with PCA}

We created energy spectra every 0.125 and 0.250 sec from the Event
mode (E\_125us\_64M\_0\_1s) data of all the PCUs that were on during
the burst.
Given the high count rates during the peak of the bursts, we corrected
each energy spectrum for dead-time using the methods suggested by the
RXTE team\footnote{http://heasarc.gsfc.nasa.gov/docs/xte/recipes/pca\_deadtime.html}.
For each energy spectrum, we created the corresponding response matrix
using the latest information available on the response of the
instrument at the relevant times.
As is common practice, we used as background the energy spectrum of
the persistent emission taken seconds before each burst\footnote{We
  used 100 sec of the persistent emission to calculate the spectrum.
  However, we found no significant differences in the fits when the
  persistent-emission before or after the burst was used, or when
  using data-segments of different lengths -between 100 and 500
  seconds.}.
We used a black-body model to fit the resulting burst spectra; this
choice allows us to compare our results with previous work on burst
spectra in other sources.
With the procedure described above we are assuming that the X-ray
spectra, after the persistent emission has been subtracted, are
Planckian and that the observed luminosity of the source is:

\begin{displaymath}L = 4 \pi \sigma T^4 R^2, \end{displaymath}

\noindent so the unabsorbed bolometric X-ray flux is determined using

\begin{displaymath}F_{bol} = \sigma T^4 (R/D)^2, \end{displaymath} 
where $\sigma$, \textit{T}, \textit{R} and \textit{D} are the
Stefan-Boltzmann constant, the black-body temperature, the neutron
star photosphere radius, and the distance to the source, respectively.
The ratio $(R/D)^2$ is the normalization of the black-body model ({\tt
  bbodyrad} -- see XSPEC manual for details).
We note that from the ($\chi^2$) statistical point of view, the
spectra of X-ray bursts are generally well described by black-body
emission. However, the emission from the neutron star and its
environment (e.g., accretion disk) is expected to be more complex than
simple black-body emission \citep[see, e.g., ][ and references
therein]{Vanparadijs82,London84,Kuulkers03} and T is a color
temperature.

\begin{figure}
\centering
\resizebox{1\columnwidth}{!}{\rotatebox{0}{\includegraphics{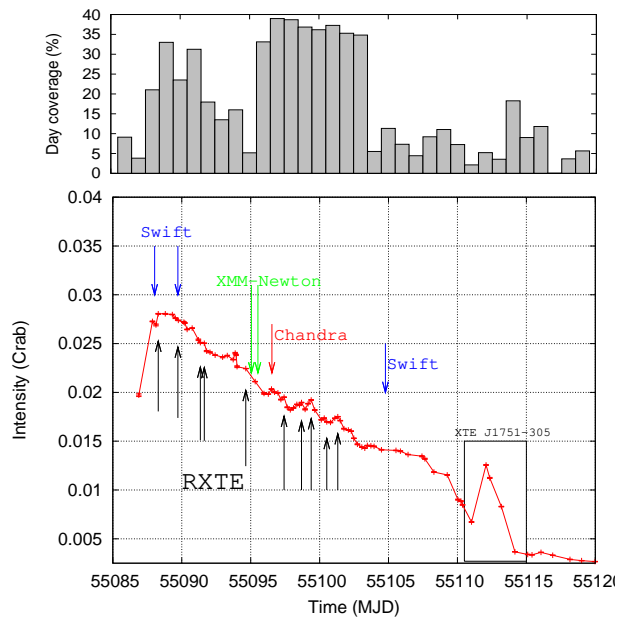}}}
\caption{\textit{Lower panel:} Intensity (2-16 keV) normalized by the
  Crab. With blue, black, red and green arrows we mark when X-ray
  bursts were detected by Swift, RXTE, Chandra and XMM-Newton
  respectively. The flare at the end of the outburst corresponds to
  the outburst of a nearby AMXP XTE~J1751--305 \citep{Markwardt09} in
  the PCA field of view. 
  \textit{Upper panel:} Percentage of the day covered by RXTE
  data. This estimates are based on Standard 2 data.}
\label{fig:lc}
\end{figure}

\begin{figure} 
\centering
\resizebox{1\columnwidth}{!}{\rotatebox{-90}{\includegraphics{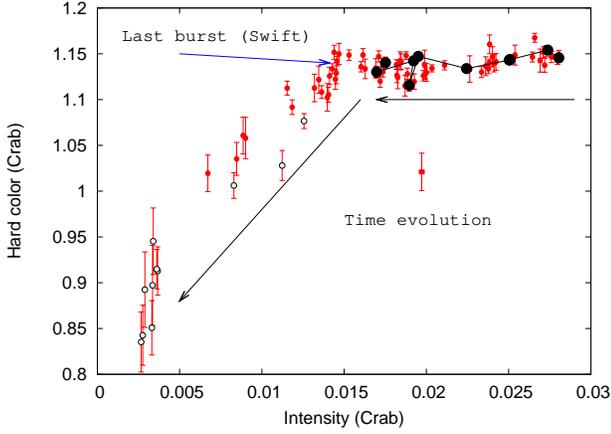}}}
\caption{Hardness-intensity diagram normalized by the Crab. Black
  arrows show how the spectral characteristics evolve in time. The
  blue arrow marks the time when the last X-ray burst was detected (in
  a Swift observation). Black filled circles mark the times when X-ray
  bursts were detected with RXTE. Empty circles mark observations
  where the AMXP XTE~J1751--305 was also active.}
\label{fig:hid}
\end{figure}

\subsubsection{Burst oscillations}\label{sec:BOs}

Burst oscillation analysis was conducted using 125 $\mu$s time
resolution PCA event mode data, barycentered using the refined
position \citep[$\alpha=17^h 51^m 08^s.66$, $\delta = -30^o 57^{'}
41^{''}$, see][]{Nowak09}.
We looked for data overruns which sometimes occur during bright
bursts, but found none.  We then extracted data in the 2-25 keV range
where the bursts are dominant.

Burst oscillation fractional rms amplitudes were computed using the
procedure outlined in Section 2.1 of \citet{Watts05}.  
We measure the Z$^2$ statistic, an unbinned version of the standard
power spectrum.  From this the best estimate of the pure source-only
signal power in the absence of measurement noise,
$Z^2_{\mathrm{signal}} = Z_s$, is taken to be that for which the
cumulative probability $f_n$ of obtaining the measured power
$Z^2_\mathrm{measured} = Z_m$, $f_n(Z_s:Z_m) = 0.5$.
Errors on $Z_s$ are given by the points where $f_n(Z_s:Z_m)$ is 0.159
and 0.841 (equivalent to $\pm 1 \sigma$ for a normal
distribution). The fractional rms amplitude $r$ is then given by

\begin{equation}
r = \left(\frac{Z_s}{N}\right)^{1/2}\left(\frac{N}{N-N_b}\right)
\end{equation}
where $N$ is the total number of photons, and $N_b$ is the number of
background photons.

We first computed burst average fractional rms amplitude, assuming a
constant frequency, for the entirety of each burst.  To reduce the
contribution from the accretion-powered pulsations (which will
contribute to the pulsed fraction if accretion continues unaffected
during the burst), we used data from the period where the burst flux
exceeds twice the pre-burst level. If the accretion flow and the
fractional rms amplitude of the persistent pulsations remain unchanged
during the burst, then the magnitude of the correction is given by

\begin{equation}
r = \frac{r_\mathrm{bur}N_\mathrm{bur} +
  r_\mathrm{acc}N_\mathrm{acc}}{N_s}
\label{contamination}
\end{equation}
where $N_\mathrm{bur}$ and $N_\mathrm{acc}$ are the number of source
photons from the burst and accretion processes respectively, with
$r_\mathrm{bur}$ and $r_\mathrm{acc}$ being the fractional rms
amplitudes of the two different processes.  The total number of source
photons is $N_s = N_\mathrm{bur} + N_\mathrm{acc}$.

The average fractional rms amplitude for each burst, computed in this
way, is given in Table~\ref{table:bursts}.  We find significant drifts
in the frequency of the burst oscillations during many of the bursts:
the amplitude that we report is for the constant frequency where the
$Z^2$ statistic is maximized.
Fitting the frequency drifts would slightly increase the measured
amplitude.
Burst average fractional rms amplitudes for the first harmonic were
also computed.

To monitor the variation in frequency and amplitude over the course of
each burst, we computed dynamical power spectra using overlapping
4s windows, to detect any frequency drifts.
We also calculated fractional rms amplitudes for overlapping segments
of 5000 photons (to ensure comparable statistics throughout the
bursts). This is the same procedure as that adopted in
\citet{Watts05}.  Amplitudes are only calculated for bins where the
measured power exceeds 14 (equivalent to a 3$\sigma$ single trial
detection).

\section{RXTE Results}\label{sec:results}

\begin{figure} 
\centering
\resizebox{1\columnwidth}{!}{\rotatebox{0}{\includegraphics{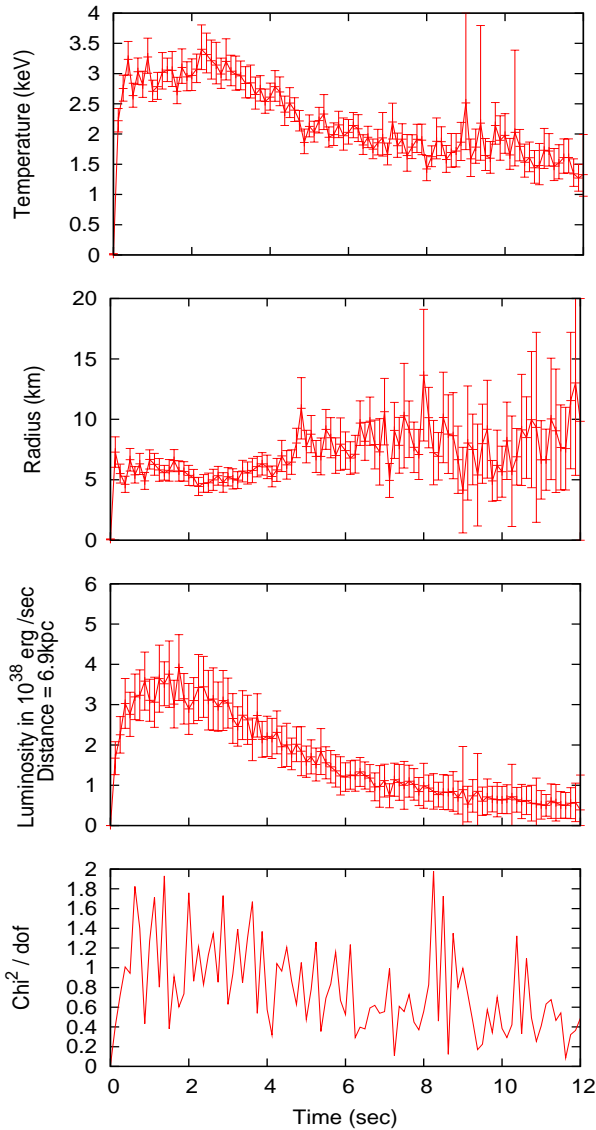}}}
\caption{Temperature, radius and luminosity evolution for the brightest
  burst in our sample (Burst 9, see Table~\ref{table:bursts}). In the
  lowest panel we show the $\chi^2/dof$ of our fits. The Radius and
  Luminosity were calculated assuming a distance of 6.9 kpc}
\label{fig:evolution}
\end{figure}

\begin{figure} 
\centering
\resizebox{1\columnwidth}{!}{\rotatebox{0}{\includegraphics{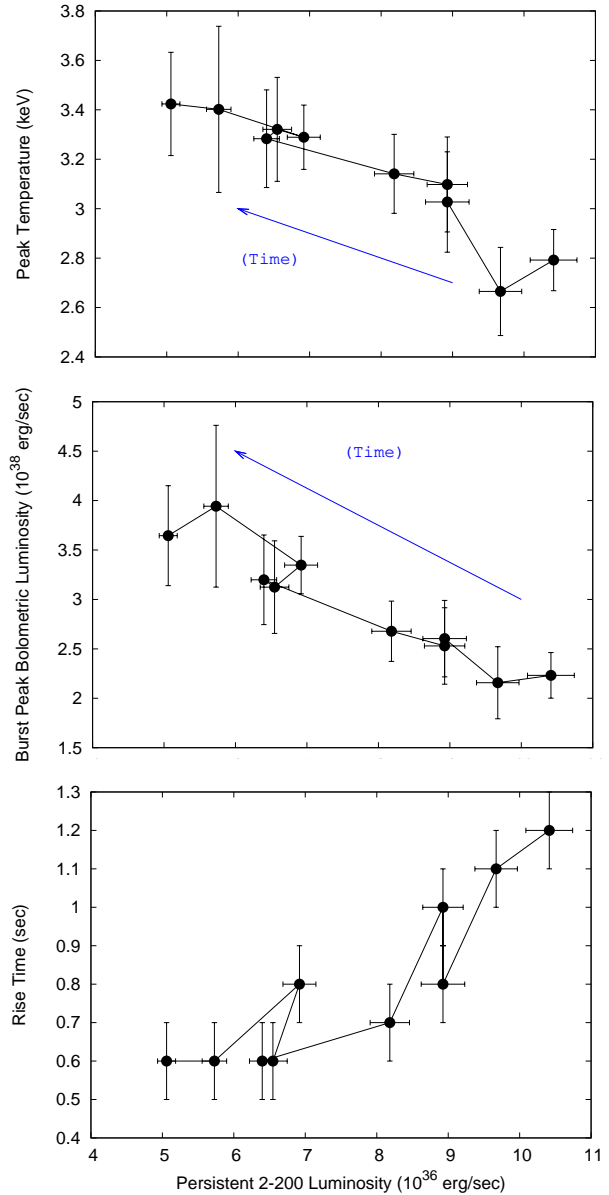}}}
\caption{Burst peak black body temperature (upper panel), bolometric
  luminosity at the peak (middle panel) and burst rise time (lower
  panel) as a function as of the 2-200 keV luminosity of the source
  before the occurrence of the X-ray bursts. To estimate the
  luminosity we assumed a distance of 6.9~kpc. Rise times were
  calculated using 0.1 s resolution.}
\label{fig:temperature}
\end{figure}

\subsection{The outburst evolution and the occurrence of X-ray
  bursts}\label{sec:lc}

Figure~\ref{fig:lc} shows the 2--16 keV light curve of
IGR~J17511--3057. The outburst was detected on September 12th, 2009
(MJD 55086) by INTEGRAL and by RXTE. On this day IGR~J17511--3057 was
already about 20 mCrab; its intensity increased for about 2 more days,
reaching a peak luminosity of $\approx27$~mCrab (corresponding to a
$\approx$4.5\% (d/6.9)$^2$ Eddington luminosity as estimated from
spectral fits in the 2--200 keV range). After that, the intensity
decreased almost linearly until MJD $\approx55107.5$, when the
intensity appears to decrease faster in time.
The weakening of the source was interrupted by an apparent flare which
started on MJD $\approx55111$ and lasted for about 3 days. This
flare-like increase of intensity was not from IGR~J17511--3057, but
due to the nearby AMXP XTE~J1751-304 \citep{Markwardt09a} in the PCA
field of view.

In Figure~\ref{fig:lc} we mark the episodes of X-ray bursts as
detected by RXTE (10 bursts), Swift \citep[3 bursts, see ][]{Bozzo10},
Chandra \citep[1 burst, see][]{Nowak09} and XMM-Newton \citep[2
bursts, see][]{Papitto09a,Papitto10}.
After MJD $\approx55105$ no other bursts were detected by any
satellite, suggesting a drop in burst rate. However, note that this
drop in burst rate also coincides with the moment when the PCA
sampling decreased significantly (see upper panel of
Figure~\ref{fig:lc}).
In Figure~\ref{fig:hid} we plot the hardness--intensity diagram using
all PCA available data of IGR~J17511--3057. At the beginning of the
outburst decay, the hard color remains constant as the intensity
decreases. A break occurs when the intensity reaches $\approx15$ mCrab;
the hard color then starts to decrease approximately linearly with
intensity.

\subsection{X-ray burst spectral characteristics}\label{sec:burstspectra}

We found 10 Type I X-ray bursts in the PCA data;
all of them were well fitted with a black-body model (with average
$\chi^2$/dof between 0.8 and 1.2 during the first 5-10 seconds of the
burst).
All bursts showed similar temperature, radius and bolometric flux
profiles (see Table~\ref{table:bursts}). The temperature and flux
profiles were all single peaked, with no evidence for precursors.
All bursts reached their maxima within 1.2 seconds.
The rise times \citep[from 25\% to 90\% of peak flux, as defined in
][]{Galloway08} of all bursts were between 0.6 to 1.2 sec.
The maximum blackbody temperature (\textit{kT}) was always between 2.5
and 3.5 keV and peak bolometric fluxes were in the $(3-6.7) \times
10^{-8}$ erg s$^{-1}$ cm$^{-2}$ range.
In all cases, the black-body radius remained approximately constant
for the first $\approx3$ seconds of the burst, then increased until
reaching an average value of $\approx9$ km (assuming a distance of
6.9~kpc).  
All bursts were also similar as far as decay shape. Unlike other cases
where double-exponential or power law decays are favored, a simple
exponential gives a statistically good fit in most cases.  Decay times
were in the range 5-8s. 
The total energy of each burst $E_{tot}$ was always in the range
$(2.5-3) \times 10^{39} ($d$/6.9 kpc)^2$ erg.
The burst timescales $\tau$, defined as the ratio between $E_{tot}$ /
$L_{peak}$, were all in the range 5.5-9s. In
Figure~\ref{fig:evolution} we show the temperature, radius and
luminosity profiles for the brightest burst in our sample (i.e.,  Burst
9, see Table~\ref{table:bursts}).

In Figure~\ref{fig:temperature} we show the black-body temperature
(upper panel), the bolometric luminosity at the peak of each burst
(middle panel) and the burst rise time (lower panel) against the 2-200
keV luminosity of the source before the bursts, respectively.
Both the temperature and the bolometric luminosity appear to
anti-correlate with the persistent luminosity of the source; however,
there is a clear correlation between the burst rise time and the
persistent luminosity. 
To further investigate the significance of the anti-correlations, we
calculated the Spearman rank correlation coefficients; we found
coefficients of $-0.948$ and $-0.936$ for the temperature and the
burst-peak luminosity, respectively (the two-sided significance of
their deviations from zero equals $2.9 10^{-5}$ and $6.7 10^{-5}$,
respectively).
We also studied the significance of the correlation between rise time
and persistent luminosity and found Spearman rank correlation
coefficient of $+0.941$, with a $5.01 10^{-5}$ two-sided significance
of its deviations from zero.
Clearly, the apparent relations in Figure~\ref{fig:temperature} are
all significant.

\subsection{Distance to IGR J17511--3057}\label{sec:distance}

None of the bursts showed indications of photospheric radius
expansion.  By using the highest measured bolometric peak flux of $6.7
\times 10^{-8}$ erg s$^{-1}$ cm$^{-2}$ we can estimate an upper limit
on the distance.
We find a distance $D<6.9$~kpc when using the empirically determined
Eddington luminosity of $3.79 \times 10^{38}$ erg s$^{-1}$ \citep[this
$L_{Edd}$ is an average value estimated from PRE bursts from sources
situated in globular clusters, see ][]{Kuulkers03}.
This is consistent with the estimations from XMM-Newton data recently
reported by \citet{Papitto10}.
Using the same model as \citet{Galloway08}\footnote{The approximation
  was recently used by \citet{Galloway08} to compare a sample of more
  than a thousand X-ray burst from different sources. The distance is
  given by: \vspace{0.5cm}

\begin{math}
D = 8.6 \cdot 
(\frac{\mathrm{Flux}_{\mathrm{Bol}}}{3\cdot10^{-8} \mathrm{erg} \ \mathrm{cm}^{-2} \ \mathrm{s}^{-1}})^{-1/2} \cdot
(\frac{\mathrm{M}_{\mathrm{NS}}}{1.4\mathrm{M}_{\odot}})^{1/2} \cdot \\
~~~~~~~~~~~ \cdot (\frac{1+\mathrm{z(R)}}{1.31})^{-1/2} \cdot
(1+\mathrm{X})^{-1/2} \mathrm{kpc}
\end{math}
\vspace{0.5cm}

where $M_{NS}$ is the mass of the neutron star in solar masses, $X$ is
the mass fraction of hydrogen in the neutron star atmosphere and z(R)
is the term that takes into account the gravitational redshift at the
photosphere \citep[were $1+z(R) = (1-2GM_{NS}/Rc^2)^{-1/2}$, $G$ is
the gravitational constant, $c$ the speed of light and $R$ the radius
measured at the photosphere -- see][]{Galloway08}. }
and standard values for the mass and the radius of the neutron star
(i.e. $M_{NS}=1.4M_{\odot}$ and $R=10$~km), we found $D<5.76$~kpc and
$D<4.4$~kpc for hydrogen mass fractions of $X=0$ and $X=0.7$,
respectively. Higher values of the radius and the mass give higher
upper limits (at the very conservative case of $M_{NS}=2.2M_{\odot}$,
$R=15$~km and $X=0$, we find $D<7.1$~kpc).

\subsection{Burst oscillations}

\begin{figure*}
\centering
\includegraphics[width=8cm]{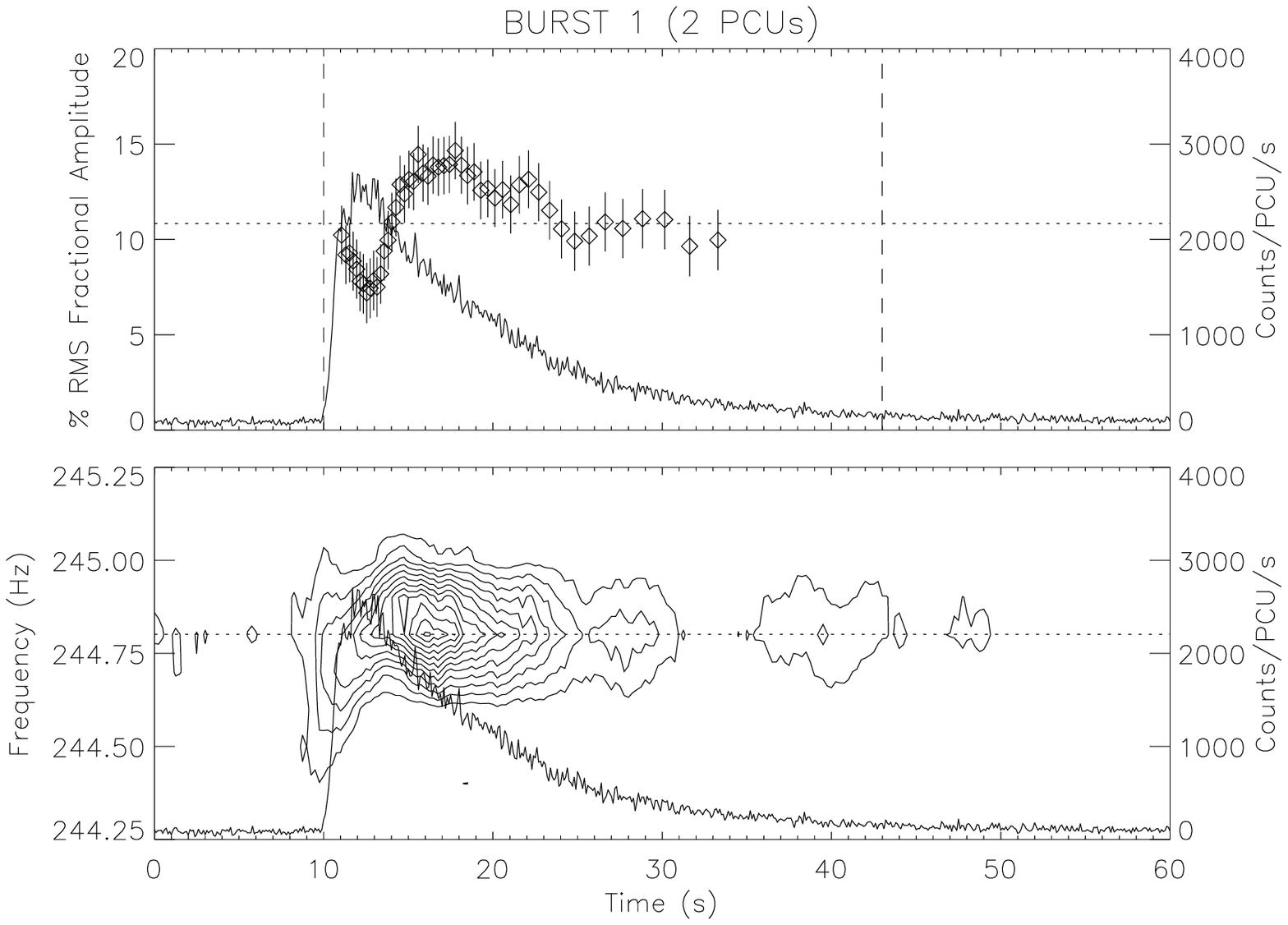}\hspace{0.5cm}
\includegraphics[width=8cm]{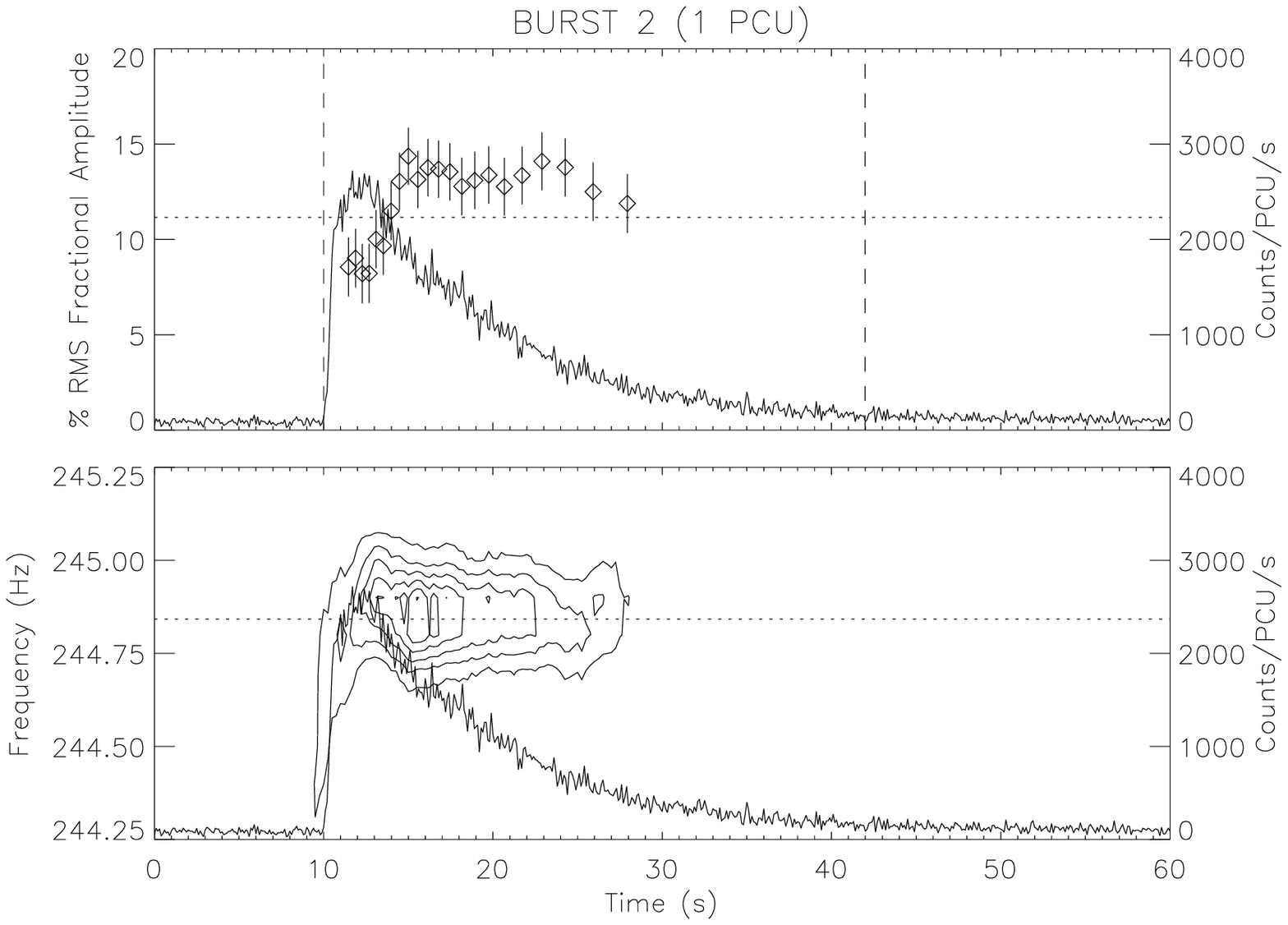}\vspace{0.5cm}
\includegraphics[width=8cm]{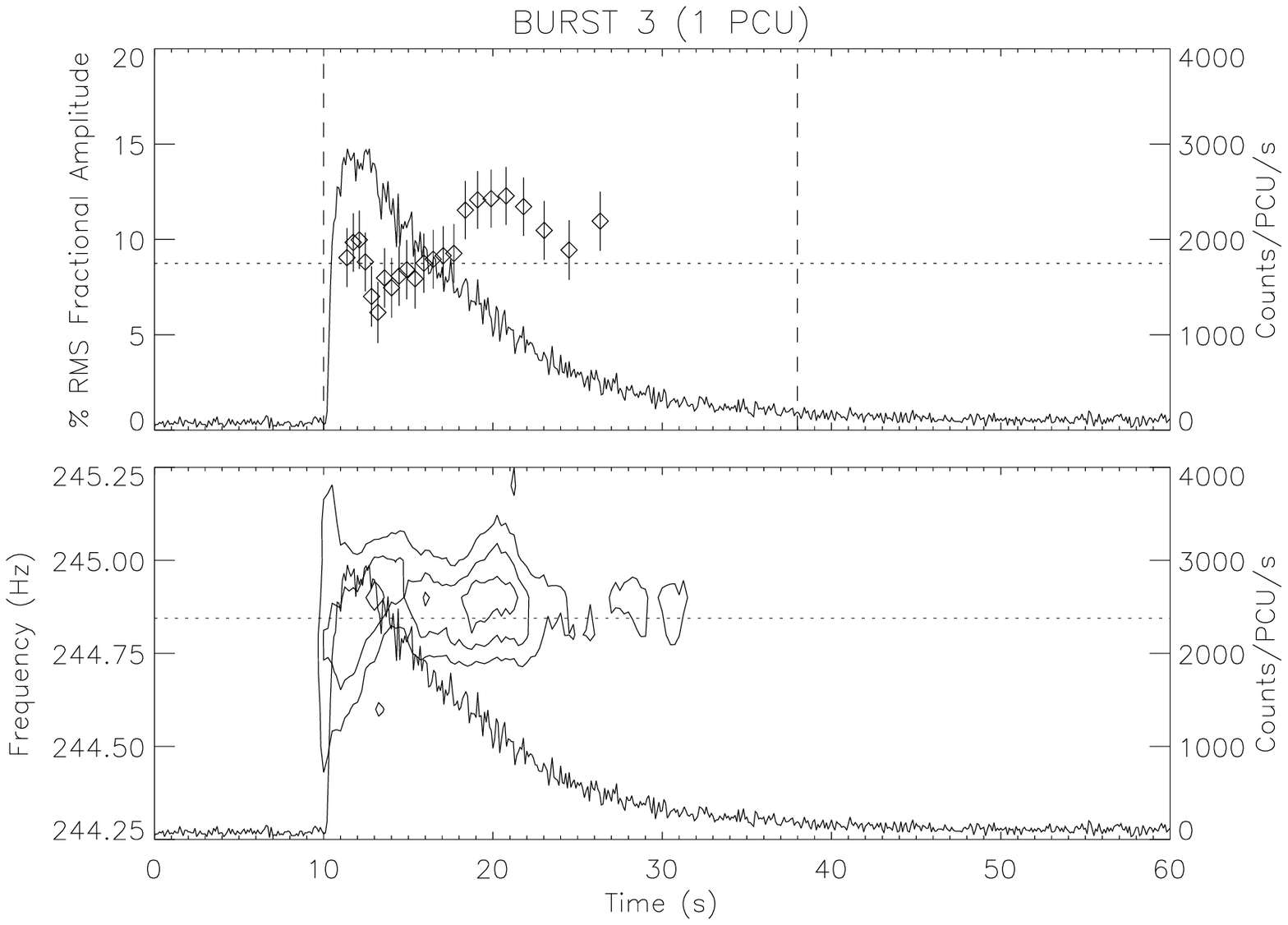}\hspace{0.5cm}
\includegraphics[width=8cm]{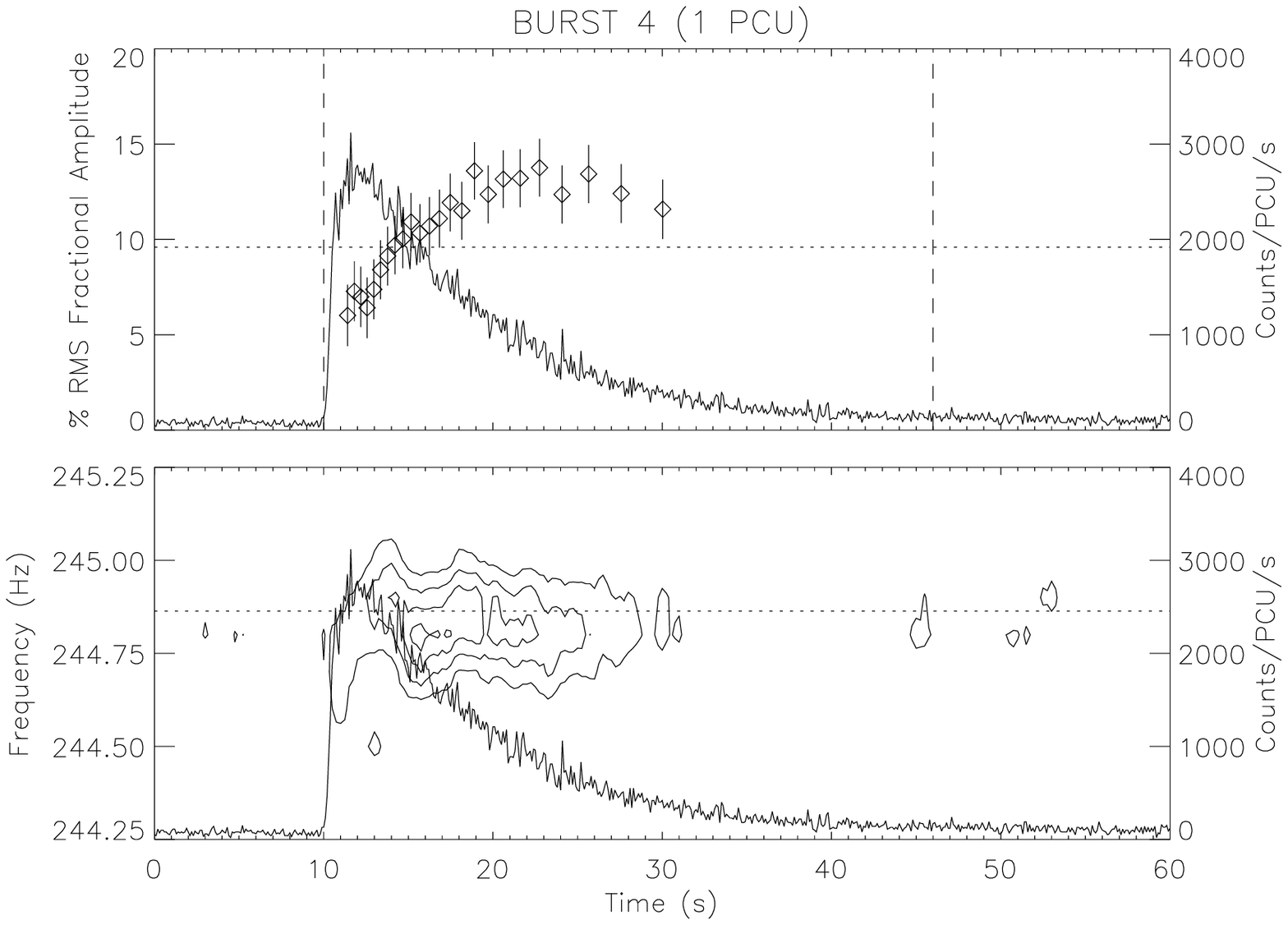}\vspace{0.5cm}
\includegraphics[width=8cm]{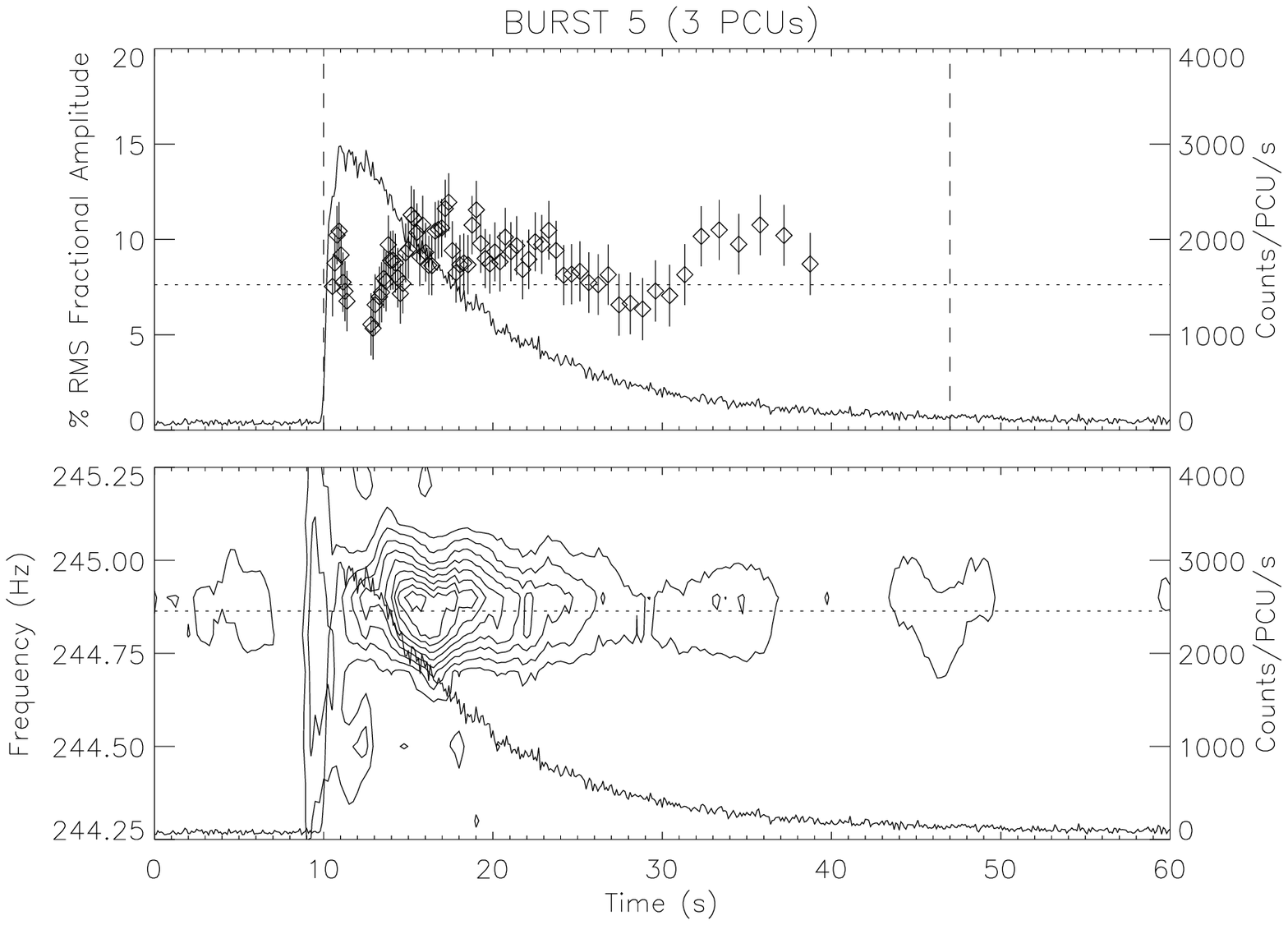}\hspace{0.5cm}
\includegraphics[width=8cm]{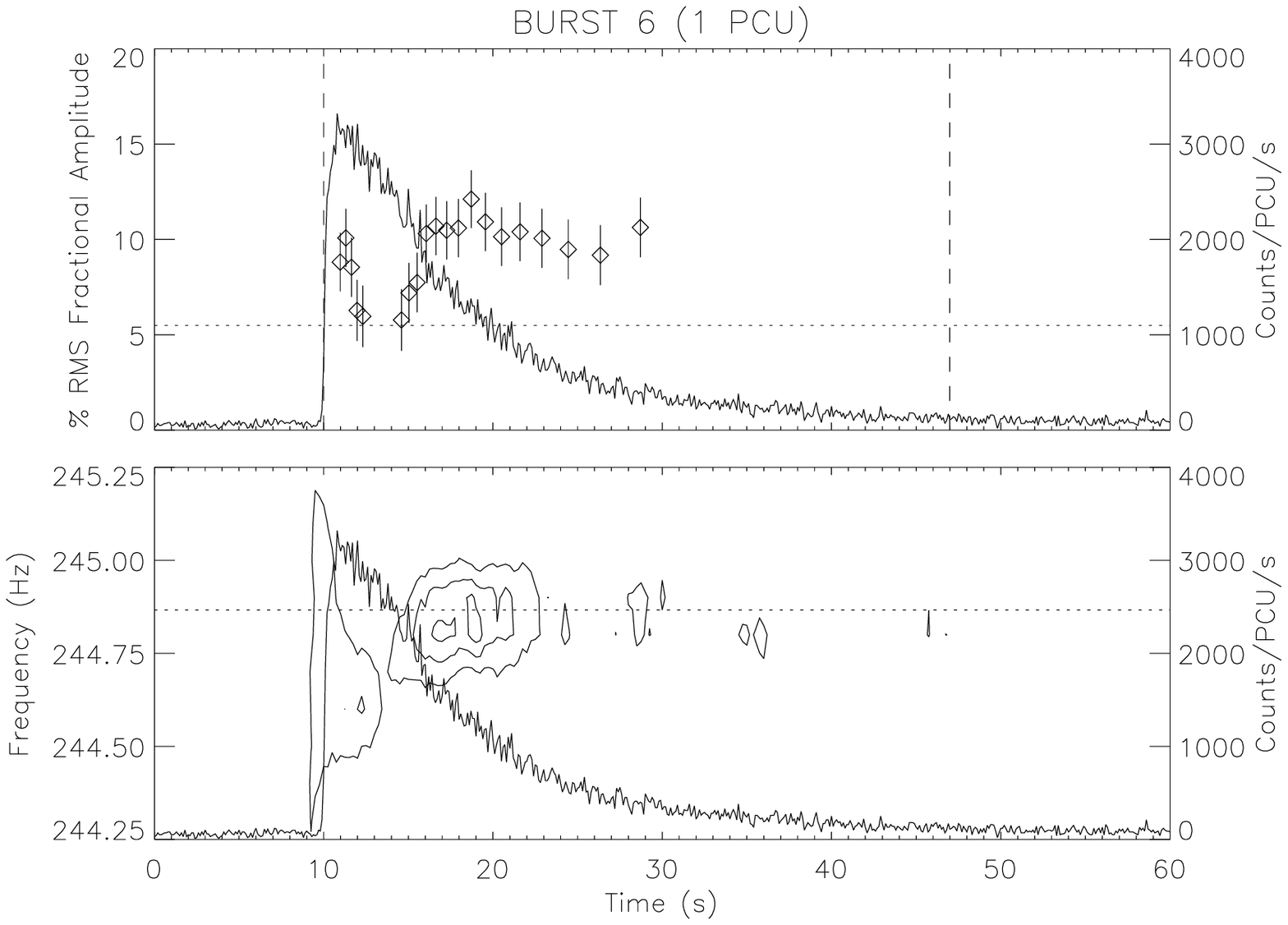}\vspace{0.5cm}
\caption{Variation of fractional rms amplitude and frequency of burst
  oscillations during each burst (2-25 keV) - plot continued overleaf.
  Scales are identical in each panel for ease of comparison.  Burst
  lightcurves are shown in each panel.  Top panels: \% RMS fractional
  amplitude computed for bins of 5000 photons, overlapped by 1000
  photons, where the countrate is at least twice the pre-burst rate.
  Amplitudes are only shown where the power constitutes at least a
  $3\sigma$ (single trial) significance.  The dotted line shows the
  integrated amplitude computed assuming a constant frequency across
  the entire period (as reported in Table~\ref{table:bursts}.  Lower
  panels: Dynamical power spectra computed for bins of 4s, overlapped
  by 0.25s.  Minimum power contour 20, increasing in increments of
  20. The dotted line marks the spin frequency of IGR~J17511--3057 at
  the beginning of the X-ray burst. }\label{fig:indivbursts}
\end{figure*}

Burst oscillations are detected significantly in all bursts observed
from this source, at a frequency very close to the known spin
frequency.
The properties of the burst oscillations in IGR~J17511--3057 are
summarized in Table~\ref{table:bursts}.
In Figure~\ref{fig:indivbursts} we show the fractional rms amplitude
(upper panels) and the dynamical power spectra for each burst (lower
panels) as a function of time.


Burst oscillations are detected throughout the bursts only for the
weaker bursts seen in the earlier part of the outburst. As the
intensity (and probably accretion rate) drops and the bursts become
brighter, and the characteristics of the burst oscillations change.
Burst oscillation amplitude drops below the detectability threshold
first in the peak of the bursts, and then also in the rise.
The fractional rms amplitude of the burst oscillations, when detected,
lies between 5 and 15\%.  Amplitude tends to be stronger in the tail
than during the rise and peak.  The burst-integrated fractional
amplitude (see dotted horizontal line in the top panels of
Figure~\ref{fig:indivbursts}) drops as the persistent luminosity
decreases and the X-ray bursts become stronger.  This drop can be
attributed to the fact that the oscillations are no longer detected
throughout the burst.  Harmonic content is detected during some bursts
(see Table~\ref{table:bursts}); this is consistent with what has been
found for the AMXP XTE J1814-338 \citep{Strohmayer03a,Watts05}

The burst oscillations show evidence for frequency drifts of $\approx
0.1$ Hz in the rise.  The frequency tends to start below the spin
frequency, reach a maximum in the peak and then stabilize very close
to the spin frequency in the tail.  In some cases the burst
oscillation frequency in the tail appears to be higher than the spin
frequency (see for example bursts 3, 9 and 10), however this is
potentially within the error bars on our contours and also within
possible errors on spin frequency due to the timing noise.  A detailed
analysis of the frequency drifts and the relation to the spin
frequency (phase-locking), including any evidence for overshoot of the
spin frequency \citep[as observed for SAX~J1808.4-3658,
see][]{Chakrabarty03} will be presented in a companion paper (Patruno
et al. in prep).

\begin{figure*}
\includegraphics[width=8cm]{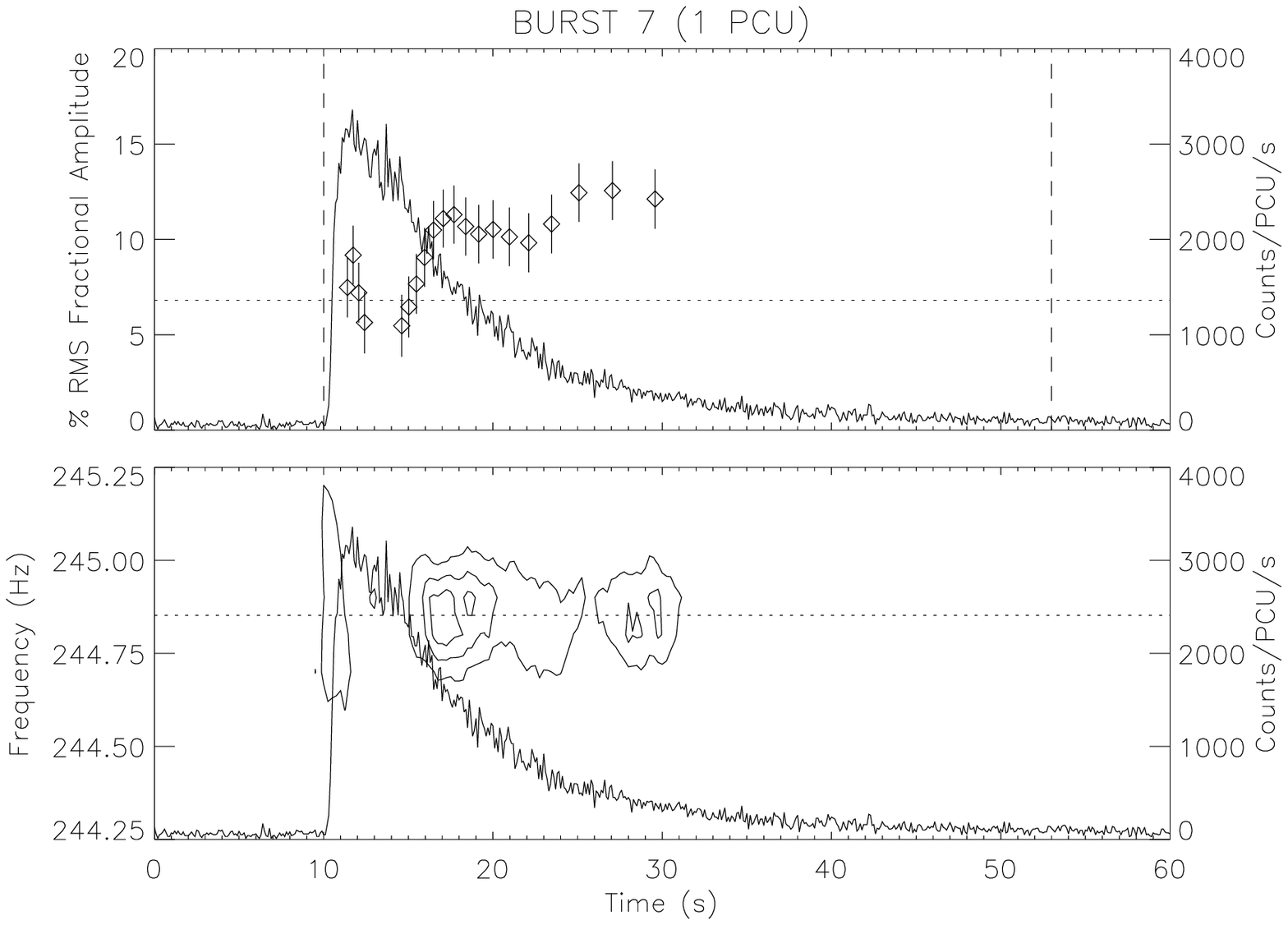}\hspace{0.5cm}
\includegraphics[width=8cm]{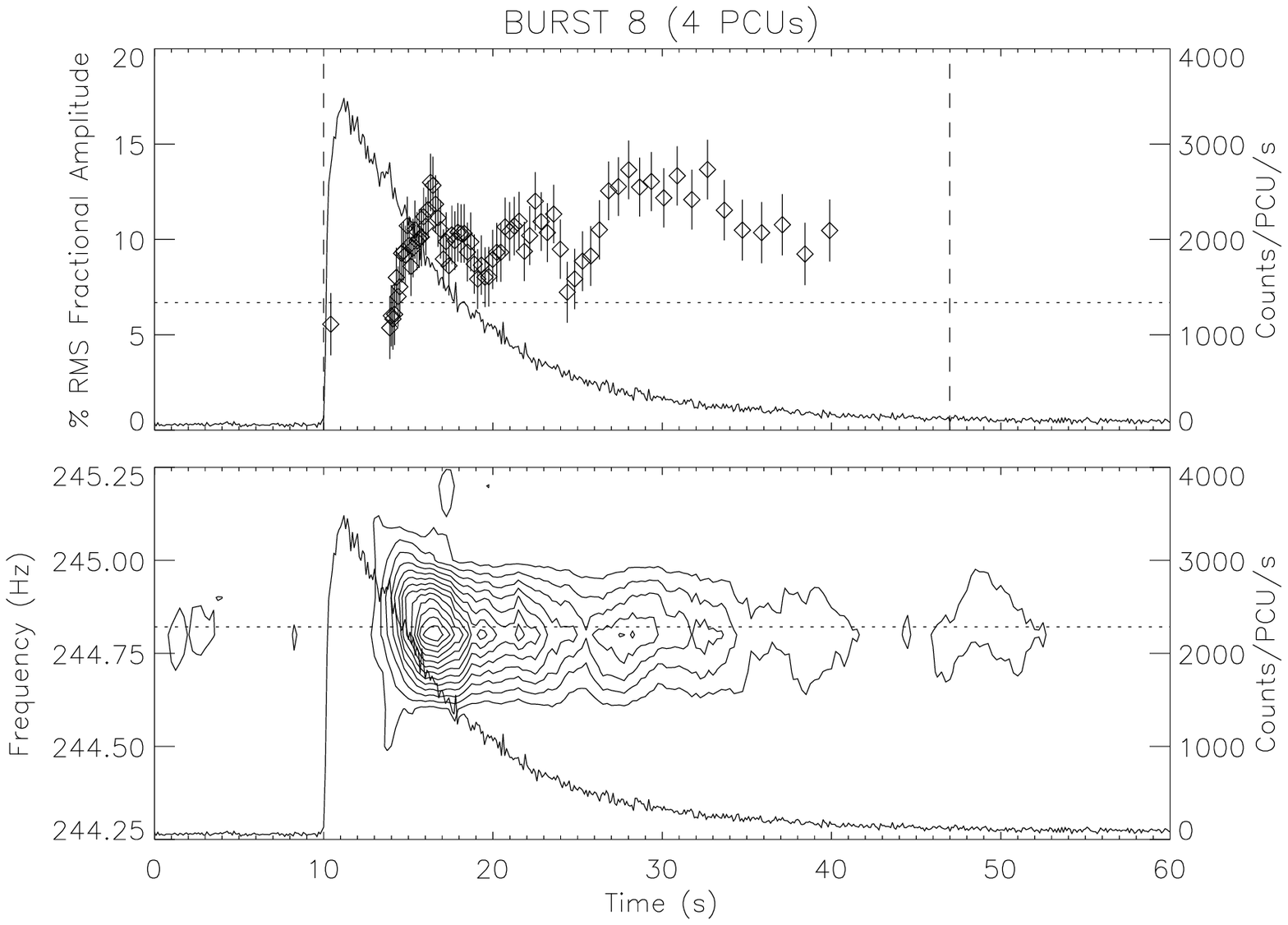}\vspace{0.5cm}
\includegraphics[width=8cm]{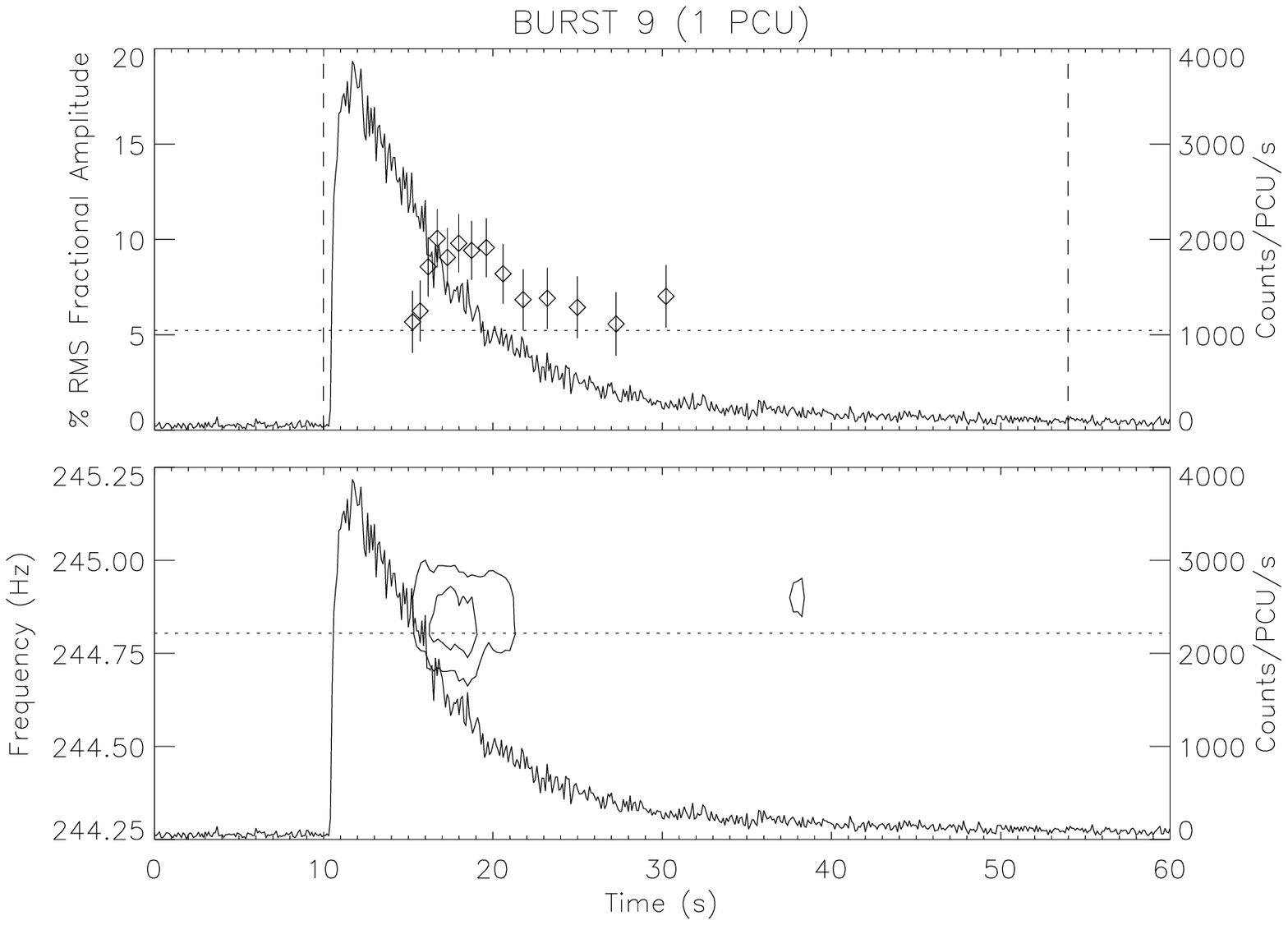}\hspace{0.5cm}
\includegraphics[width=8cm]{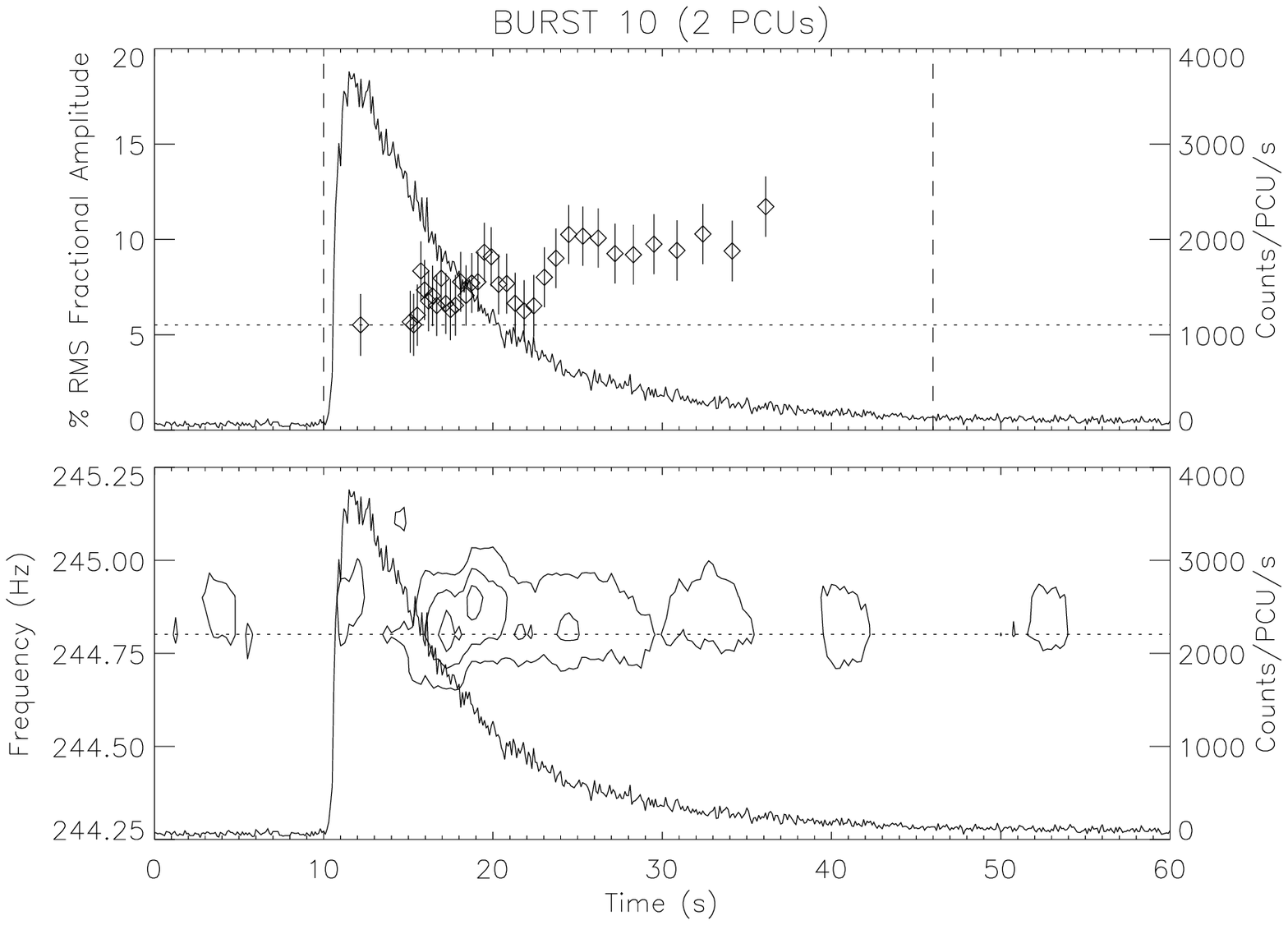}\\
Figure~\ref{fig:indivbursts} continued. 
\end{figure*}

\section{Swift XRT/BAT results}

\subsection{Upper limits on burst oscillations in the XRT data}

Of the 3 Swift bursts \citep[see, e.g., ][]{Bozzo10}, we examined
bursts 1 and 2, which occurred at UTC times 2009-09-14T00:50:27 and
2009-09-15T17:17:19, respectively.  Swift XRT Windowed Timing (WT)
mode are available for those bursts, with a time stamp resolution of
1.766 ms (Nyquist-limited sampling frequency of 283.126 Hz). We found
no BAT "failed" triggers that coincide with these X-ray burst, i.e.,
only data from XRT instrument is available for these bursts.
In the 1--7 keV band, the source peak count rate is $\approx$140 ct/s
during both bursts, with a pre-burst background of $\approx$8.5 ct/s.  We
examined FFT power spectra with 2 s durations from T-10 s to T+50 s
around the burst epochs listed above.  No oscillations were
detectable.  At the pulse frequency, the 95\% upper limit is 14\% rms
during the peak of the burst.  The upper limit is less constraining in
the tail of the burst, with typical values of 20--25\%.

\subsection{Upper limits on burst oscillations in the BAT data}

We examined the Swift BAT data for the third burst, which was reported
to occur at time 2009-09-30T18:31:57 \citep[UTC, see ][]{Bozzo10}.  We
searched for oscillations in the BAT event data, which has a time
stamp resolution of 100 $\mu$s.  In a light curve with 1 s bins, we
found a significantly enhanced BAT count rate in the time range T-3 to
T+6 compared to the reported burst time.  Since BAT is typically a
background-dominated instrument, we attempted to maximize the signal
to background ratio in order to maximize sensitivity to
oscillations. Automatic background subtraction using the BAT mask
weighting technique reduces the effective area by about 44\%.
Instead, we used the raw events and use bracketing 5 s background
intervals on both sides of the burst.  We found that the energy range
12--18 keV maximizes the signal to background ratio for this burst.
Below about $\approx$ 12 keV, the BAT energy calibration is less well
known, and various effects like passive absorption and detector energy
thresholds introduce further complications.  Above 18 keV, there are
still detectable burst counts, but the dominant background introduces
more noise.  Finally, we removed noisy detectors, disabled detectors,
and detectors whose fractional area exposure to this source was less
than 50\%.  Detectors with less fractional exposure contribute
primarily background.  After these cuts, the background rate was 1806
ct/s in 13096 detectors (out of 32768 possible detectors).  We
constructed a single FFT power spectrum in the 9 s window described
above and detected no significant oscillations.  We set a 95\% upper
limit to oscillations at the pulse frequency of 10.4\% rms of the net
burst emission.  We also constructed individual 2 s power spectra to
search for shorter term variability at the pulse frequency, but did
not detect any.

\section{Discussion}\label{sec:discussion}

We have presented a detailed analysis of the ten X-ray bursts detected
by RXTE from the recently discovered AMXP IGR~J17511--3057.  These
bursts were recorded from the peak of the outburst onwards as
accretion rate fell by a factor $\approx 2$.  The bursts become
brighter and shorter as accretion rate falls.
This behavior is consistent with the fraction of He in the bursts
increasing as the accretion rate falls, as H has more time to burn to
He via the hot CNO cycle prior to ignition \citep[consistent with the
$<$4\% Eddington luminosity/accretion rate we observe.  See,
e.g.,][]{Bildsten98,Bildsten00}.

Burst oscillations were detected in all of the bursts from
IGR~J17511--3057 recorded by RXTE.  In this sense IGR~J17511--3057 is
like the other two persistent AMXPs and unlike both the intermittent
AMXPs and the non-pulsars, where burst oscillations are detected in
only a subset of bursts.  This may be due to the fact that we observe
the persistent AMXPs over a smaller range of accretion rates than the
other sources; alternatively it may be associated with the persistent
presence of a channeling magnetic field.

The burst oscillations from IGR~J17511--3057 lie within $\approx 1$ Hz of
the known spin frequency, confirming the already well-established link
between burst oscillation frequency and spin frequency. 
The oscillations also exhibit frequency drifts.  We observe rapid
drift in the burst rise, starting below the spin rate and ending up
very close to it.  There is some suggestion of overshoot which will be
explored more rigorously in a companion paper.  
The burst oscillation frequency stabilizes very close to the spin
frequency early in the tail.  
This behavior is unusual compared to that seen in the intermittent and
non-pulsars, where the drift is slower and often persists throughout
the tail \citep[see, e.g.,][]{Muno02a,Watts09}. The drifts seen in
IGR~J17511--3057 are similar to those seen in the AMXP
SAX~J1808.4--3658 \citep{Chakrabarty03}; the highly stable burst
oscillations seen in XTE~J1814--338 remain unique
\citep{Strohmayer03a, Watts08a}.

Oscillations remain above the detectability threshold throughout the
bursts for the weaker bursts.  As the bursts get brighter, however,
the oscillations fall below the detectability threshold - first in the
peak, and eventually also in the rise.  In many sources (including the
AMXP SAX~J1808--3658) the vanishing of oscillations in the peak is
associated with the onset of PRE. PRE could in principle obscure or
shear out any asymmetry \citep[see, e.g.][]{Cumming05}; alternatively
convection during PRE bursts may suppress the oscillation mechanism
\citep{Cooper08}.  However this is not the case for IGR~J17511--3057,
since none of the bursts show evidence for PRE.
Another possibility is that there is strong frequency drift in the
rise and peak: this would reduce the possibility of detection since
our search assumes a constant frequency.  To test this we searched for
signals in shorter time windows than those used in
Section~\ref{sec:BOs}, but found no evidence to support this
hypothesis. 
Increased He fraction in the ignition column (as accretion rate falls
so there is more time to burn H to He via the hot CNO cycle) may be
important, but how this could affect the detectability or generation
of oscillations in the rise and peak is not clear.

The disappearance of oscillations in the peak and rise of bursts
without PRE is also seen in some non-pulsars \citep{Galloway08} but
this is the first time that it has been seen in bursts from a
persistent AMXP\footnote{\citet{Strohmayer03a} found that oscillations
  disappeared in the peak of the brightest burst from XTE J1814--338
  (which occurred at the lowest accretion rate).  This burst showed
  some evidence for PRE, however the spectral evidence was not
  conclusive.}.
This may point to a common mechanism for suppressing the development
of oscillations, other than PRE, that operates independently of
magnetic field.
However we note that PRE events lasting for less than 1-2 seconds
would not have been detectable: this time interval is shorter than the
periods of time ($\sim$ a few seconds) for which oscillations are
below the detectability threshold. To our knowledge there is no
theoretical constraint on how short PRE events could be; the
possibility that $<2$ sec PRE events might suppress the burst
oscillation mechanism for timescales of a few seconds cannot be
excluded and should be studied in more detail.

Burst oscillation amplitude reaches a maximum of about 15\% rms at the
fundamental frequency.  This is comparable to amplitudes seen in other
sources (pulsars and non-pulsars).  It is also similar to the
amplitude of the accretion-powered pulsations from this source.  In
this respect IGR~J17511--3057 behaves like the other persistent AMXPs
with burst oscillations.  Several of the burst oscillation trains from
IGR~J17511--3057 also show a detectable first harmonic.  The only
other source to show this consistently is the AMXP XTE J1814--338.
Burst oscillations from the intermittent pulsars and the non-pulsars
do not in general have detectable harmonic content (see, e.g.,
\citealt{Muno02b}, although see also \citealt{Bhattacharyya05} which
discusses the existence of harmonic content in the rising phase of
bursts from the non-pulsar 4U 1636--536).

\section{Conclusions}

The burst oscillations of IGR~J17511--3057 are, like those of the
other two persistent AMXPs with burst oscillations, atypical compared
to burst oscillations from the intermittent and non-pulsars.
Particularly notable are the frequency drifts and the fact that
several of the burst oscillation trains have detectable harmonic
content.  These differences between the AMXPs and the other sources
suggest that the presence of a channeling magnetic field is important
to the burst oscillation mechanism, either directly or by establishing
composition or temperature gradients.

IGR~J17511--3057 has however also broadened the spectrum of properties
exhibited by AMXP burst oscillations.  This AMXP is the first for
which oscillation amplitudes drop below the detectability threshold in
bright bursts, without the bursts exhibiting simultaneous PRE.  This
phenomenon is also seen in non-pulsars, and points to a mechanism for
suppressing burst oscillations - other than PRE - that is less
affected by the presence of a dynamically important magnetic field.

\textbf{Acknowledgment:} We thank M. Nowak for sharing the X-ray burst
time of occurrence in their Chandra data, respectively. AP
acknowledges support from the Netherlands Organization for Scientific
Research (NWO) Veni Fellowship. ML acknowledges support from the
Netherlands Organization for Scientific Research (NWO) Rubicon
Fellowship.


\clearpage

\setlength{\topmargin}{2in}

\begin{landscape}
\begin{table*}

\caption{RXTE thermonuclear X-ray bursts in IGRJ17511--3057} \label{table:bursts}
\centering
\scriptsize{
\begin{tabular}{|c|c|c|c|c|c|c|c|c|c|c|}
\hline
                                                 &       B1       &       B2       &       B3       &       B4       &       B5       &       B6       &       B7       &       B8       &       B9       &       B10       \\
ObsID$^a$                                        &   *-01-02      &   *-01-06      &   *-01-12      &   *-01-07      &   *-02-12      &   *-02-200      &   *-02-100      &   *-03-00      &   *-03-030      &   *-03-04      \\ 
Num. of PCUs On                                  &       2       &       1         &       1        &       1        &       3        &      1          &       1        &       4        &       1       &       2       \\
Start time \\(MJD+55088.)                        &  0.32967163    &  1.72041237    &  2.27323586    &    2.61714558  &  6.61841723    & 9.60216584    & 10.77723818    & 11.31361260       &  12.63290021       &  13.28983887   \\
Rise time (s)                                    &      $1.2\pm0.1$       &      $1.1\pm0.1$      &      $0.8\pm0.1$       &       $1.0\pm0.1$       &       $0.7\pm0.1$       &       $0.6\pm0.1$       &       $0.6\pm0.1$      &       $0.8\pm0.1$       &       $0.6\pm0.1$       &       $0.6\pm0.1$       \\
Decay timescale \\(s)                            &     $7.7\pm0.1$      &       $7.5\pm0.1$       &       $6.9\pm0.1$       &      $7.4\pm0.1$       &       $7.4\pm0.1$       &      $6.7\pm0.1$       &       $6.7\pm0.1$       &       $6.5\pm0.1$       &       $5.1\pm0.1$       &       $5.9\pm0.1$       \\
Tau\\(~s)                                    &     $8.6\pm1.8$      &       $8.5\pm2.3$       &       $7.5\pm1.9$       &      $7.8\pm2.0$       &       $7.9\pm1.7$       &      $6.9\pm1.7$       &       $6.8\pm1.7$       &       $6.6\pm1.2$       &       $5.5\pm1.7$       &       $6.1\pm1.4$       \\
Peak count/rate$^b$                                         &  $\sim5210$       &  $\sim2600$     &  $\sim2950$  &  $\sim2880$    &  $\sim9330$   &  $\sim3250$  &  $\sim3300$    &  $\sim15000$     &  $\sim4000$     &  $\sim7950$    \\      
 Peak Luminosity$^c$ \\($10^{38}$~erg~s$^{-1}$~(d/6.9kpc)$^2$)    &  $2.23\pm0.22$       &  $2.15\pm0.36$   &  $2.61\pm0.38$    & $2.53\pm0.38$     &  $2.67\pm0.30$    &  $3.19\pm0.45$  &  $\sim3.12\pm0.46$   &  $3.34\pm0.29$       &  $3.94\pm0.08$      &  $\sim3.64\pm0.50$    \\
 Total energy released$^c$ \\($10^{39}$~erg)         &  $1.91\pm0.38$       &  $1.83\pm0.47$       &  $1.95\pm0.46$  & $1.97\pm0.48$     &  $2.10\pm0.41$    &  $2.19\pm0.50$  &  $2.12\pm0.0.47$ &  $\sim2.22\pm0.36$       &  $\sim2.14\pm0.62$      &  $2.23\pm0.46$    \\
Duration \\(s)                                      &   $\sim38$       &      $\sim26$        &       $\sim23$        &       $\sim25$        &       $\sim30$        &       $\sim25$        &       $\sim23$        &       $\sim30$        &       $\sim17$        &       $\sim23$        \\
Persistent $L_x$ \\($10^{37}$~erg~s$^{-1}$~(d/6.9kpc)$^2$) &       $1.04\pm0.03$      &       $0.97\pm0.03$        &       $0.89\pm0.03$        &      $0.89\pm0.03$        &       $0.81\pm0.03$        &      $0.86\pm0.02$        &      $0.64\pm0.01$        &       $0.69\pm0.02$        &       $0.57\pm0.01$        &       $0.50\pm0.01$        \\
\hline
Osc. rms amplitude$^e$ \\(fundamental)        &       10.8 $\pm$ 0.4 &       11.2 $\pm$ 0.6        &       8.7 $\pm$ 0.6       &       9.6 $\pm$ 0.6       &       7.6 $\pm$ 0.3       & 5.5$^{+0.6}_{-0.5}$       &      6.8 $\pm$ 0.6       &       6.7$\pm$ 0.3 &       5.2$^{+0.6}_{-0.5}$       &       5.5 $\pm$ 0.4       \\
Osc. rms amplitude$^e$ \\ (1st harmonic)                            & 2.3$^{+0.5}_{-0.4}$       &   2.2$^{+0.7}_{-0.5}$   &       $<3.8$ &   $<3.7$    &       $<2.1$ &    $<3.6$ &      $<3.7$      &   $1.3 \pm 0.3$   &       2.3$^{+0.6}_{-0.5}$     & $<2.5$     \\
$N_\mathrm{acc}/N_\mathrm{bur}$$^g$                            &       0.09       &       0.09       &       0.08       &       0.08       &       0.08       &       0.06       &       0.07       &       0.07       &       0.06       &       0.06       \\
\end{tabular}}\\
\flushleft
\vspace{0.2cm}
$^{a}$:'*' stands for 94041-01.\\
$^{b}$:The persistent emission has been subtracted. The count rates are not normalized by the number of PCUs on. The 2.5-24 keV range was used. \\ 
$^{c}$:Bolometric (blackbody) peak Luminosity assuming a distance of 6.9 kpc.  \\ 
$^d$: Frequency given is that for which the burst average amplitude is
maximized assuming a constant frequency model (see text for details). \\
$^{e}$:Burst average fractional rms amplitude (in units of \%) of the oscillations in the 2--25 keV range
assuming a constant frequency (see text for details).\\
$^f$: RMS amplitudes are quoted when measured power
exceeded 14 (a 3$\sigma$ detection).  Where measured power is lower than this
we quote $3\sigma$ upper limits on amplitude.  Again these
measurements assume a constant frequency model.  
$^g$:  The ratio of accretion to burst photons during this period (see
Equation \ref{contamination}, corrected for background.).
\end{table*}

\end{landscape}

\end{document}